# Radiation gasdynamics of planetary nebulae – VI. The evolution of aspherical planetary nebulae

Garrelt Mellema

*Astrophysics Group, Department of Mathematics, UMIST, P.O. Box 88, Manchester M60 1QD*

**ABSTRACT**

This paper reports the results of the numerical study of the formation of aspherical Planetary Nebulae through the Generalised Interacting Winds Model, taking into account the effects caused by the evolving central star and fast wind. The results show for the first time that aspherical nebulae do form within the required time scale. Consideration of the development of the nebula shows that in the early stages it is the ionization of the aspherical AGB wind that contributes considerably to the shaping of the nebula. Furthermore the passing through of the ionization front may modifies the density distribution in the slow wind, leading to the formation of a surrounding envelope, and sometimes a different morphology for the nebula than was to be expected from the initial conditions. I consider how the different phases of ionization fronts and wind swept bubbles can be observationally distinguished.

**Key words:** Hydrodynamics – Methods: numerical – ISM: bubbles– Planetary nebulae: general

## 1 INTRODUCTION

Planetary nebulae (PNe) form around low to intermediate mass stars (main sequence mass less than 8 $M_\odot$) towards the end of their life. The nebula forms from the interaction between the slow wind lost by the star as a red giant on the Asymptotic Giant Branch (AGB) and the faster wind that follows after the AGB phase. Typical parameters for the AGB slow wind are mass loss rates of $10^{-7}$ – $10^{-4}$ $M_\odot$ yr$^{-1}$ and velocities of 5 – 25 km s$^{-1}$ (see the review by Habing 1990). After the AGB the star evolves from a cool star ($\sim$ 5000 K) into a hot star ($\sim$ 150 000 K), producing increasing amounts of ionizing photons and an accelerating fast wind (see Schönberner 1993 for a review on post-AGB evolution). The velocity of this fast wind increases from about 100 km s$^{-1}$ to 4000 km s$^{-1}$ at mass loss rates of $10^{-9}$ – $10^{-7}$ $M_\odot$ yr$^{-1}$ (see the review by Perinotto 1993).

Extensive work on the interaction between the slow and fast wind has shown that aspherical mass loss on the AGB can explain many of the observed PN morphologies and kinematics (Soker & Livio 1989; Mellema, Eulderink & Icke 1991; Icke, Balick & Frank 1992; Frank 1992; Mellema 1993; Frank & Mellema 1994b (Paper IV); Mellema & Frank 1995 (Paper V)). In all of these articles the modelling has been of the 'constant environment' type: the fast wind shaping the PN is kept constant and when radiative effects are included, the central star is considered not to evolve. This type of simulations illustrates the basics of the aspherical interacting winds model and can be used to explain the general shapes and kinematics of nebulae. However, we know that the cen-tral star is evolving on the same time scale as the nebular gasdynamics. This is also true for the fast wind, since its properties are related to the stellar parameters. So, realistic modelling of the formation of aspherical PNe should take this into account. Such 'evolving environment' models can give information on time scales and changing ionization conditions, something not contained in the constant environment models.

On the other hand, the theoretical stellar evolution tracks and the time-dependence of the fast wind are uncertain. By running complex numerical simulations based on them, one does run the risk of overinterpreting them.

In one-dimensional models of PN evolution stellar evolution tracks and changing fast wind properties have been used (Marten & Schönberner 1991; Frank 1993 (Paper II); Mellema 1994 (Paper III). They show that despite the huge uncertainties, simulations of this kind can shed light on certain aspects of PN evolution, such as the age-size relation, the dynamical influence of ionization fronts, etc.

So, although one should keep in mind the large uncertainties in the theoretical stellar evolution tracks and fast wind properties, two-dimensional models using these changing boundary conditions will certainly be of use. This paper presents the first radiation-gasdynamics simulations of aspherical PNe using such realistic boundary conditions.

One of the important questions 'evolving environment' models should help to answer is whether the various shapes the interacting winds model produces in the 'constant environment' models, can also form under more realistic conditions at the proper time scales. Keeping in mind the effects of the ionization front seen in the one-dimensional models, it is also important to see how the slow wind is modified by the varying UV flux from the central star.

This paper is organised in the following way. In Sect. 2 I briefly describe the numerical method used. Section 3 deals with initial and boundary conditions used, i.e. the prescription for the slow wind, fast wind and central star. Section 4 contains a discussion of the gasdynamic behaviour seen in the simulations. Section 5 is filled with the presentation of synthesized observations, images and long slit spectra and their comparison to real observations. In Sect. 6 the conclusions are summarized.

## 2    NUMERICAL METHOD

The interacting winds problem is difficult to tackle numerically. The numerical gasdynamics method should be of high quality, i.e. at least of second-order accuracy and with minimal numerical diffusion. The method used, the Roe solver, fulfils these requirements and has been extensively used for this problem (see the previous papers in this series). The Roe solver is an approximate Riemann solver that uses characteristics to follow the evolution of the flow. A more extensive treatment can be found in Eulderink (1993) and in Mellema et al. (1991).

Most simulations in this paper use an implementation of the Roe solver, including the exact analytical solution for the subgrid model (see Eulderink 1993, Chap. 2, Appendix C7). This version was used because the approximate method used previously often resulted in numerical problems. The exact solution is expensive, but more robust.

The accurate numerical gasdynamics method is used here together with the radiation method described in Frank & Mellema 1994a (Paper I). This method calculates the ionization structure and the heating and cooling of the gas. This is done in a time-dependent manner, which is important for the case of a ionizing source with changing properties. The method follows the ionization of H, He, N and O. Heating is by photoionization, cooling in the 'nebular' regime ($T < 10^5$ K) is by the most important forbidden lines and recombination lines, in the 'coronal' regime ($T > 10^5$ K) a standard cooling curve is used. See Paper I for more details.

## 3    INITIAL AND BOUNDARY CONDITIONS

The initial and boundary conditions are partly fixed by the grid choice and are partly based on assumptions about the physical conditions, such as the mass loss history on the AGB and the evolution of the fast wind. The spectrum of the central star can also be considered as a sort of boundary condition, supplying the ionizing photons. Most of these aspects have been discussed in the previous papers in this series and I will here only briefly summarize the choices made.

### 3.1    Grid and boundary conditions

The calculations are done on a two-dimensional grid using spherical coordinates $r$ and $\theta$ (see also Mellema et al. 1991, Sect. 3.1; Paper IV, Sect. 2). The polar angle $\theta$ runs from 0 (polar symmetry axis) to $\pi/2$ (equatorial plane). Rotational symmetry around the symmetry axis and reflection symmetry across the equatorial plane are assumed. These symmetries also determine the boundary conditions used at $\theta = 0$ and $\theta = \pi/2$.

The inner edge of the grid in the radial direction is placed at a distance $r_0$ from the origin. This inner boundary condition is supplied by the fast wind (see below). At the outer radial boundary an extrapolation based on the assumption of supersonic outflow is used (see Mellema et al. 1991, Eq. 28).

Spherical polar coordinates were chosen because they allow a simple inner boundary condition and facilitate the radiative transfer calculations. Because the grid spacing is non-uniform a slight error in the shock propagation is introduced. This error is largest at the inner boundary at about 10% but quickly drops as $1/r$ as one moves out.

### 3.2    Slow wind initial condition

At $t = 0$ the computational grid is filled with the slow wind. Its radial variation reflects the mass loss history on the AGB and its tangential variation introduces the asphericity in the evolution. Only the density varies across the grid, the velocity and temperature are assumed to be constant:

$$v_r(r, \theta) = v_0$$
$$v_\theta(r, \theta) = 0 \qquad\qquad (1)$$
$$T(r, \theta) = T_0 \, .$$

For the tangential variation of the density the same prescription as in Mellema et al. (1991) and Papers IV and V is used. The radial variation is the same as in Paper III. This means that the AGB is assumed to end with a superwind phase ($\dot{M} > 10^{-5}$ $M_\odot$ yr$^{-1}$). This superwind is smoothly joined with a lower (AGB) mass loss phase using a cosine function. See Paper III, Sect. 2.1 for a more thorough discussion of the background of this choice. Note that it is assumed that there has been no dynamical evolution in the slow wind before $t = 0$.

The precise formulation of the density variations across the grid is

$$\rho(r, \theta) = g(r) f(\theta)$$

$$g(r) = \frac{1}{2} \Big\{ (\rho_{\text{super}} + \rho_{\text{AGB}}) + (\rho_{\text{super}} - \rho_{\text{AGB}}) $$
$$\cos \Big[ \pi \min \Big( 1, $$
$$\max \Big( 0, \frac{r - (r_0 + v_0 t_{\text{super}})}{v_0 t_{\text{trans}}} \Big) \Big) \Big] $$
$$\Big\} \Big( \frac{r_0}{r} \Big)^2 \qquad (2)$$

$$f(\theta) = 1 - \alpha \left( \frac{1 - \exp(-2\beta \sin^2 \theta)}{1 - \exp(-2\beta)} \right)$$
$$\text{with } 0 \leq \alpha < 1 \text{ and } \beta > 0 \, .$$

The mass densities $\rho_{\text{super}}$ and $\rho_{\text{AGB}}$ are calculated from the mass loss rates according to

$$\rho = \frac{\dot{M}}{4 \pi r_0^2 v_0} \, , \qquad\qquad (3)$$

the times $t_{\text{super}}$ and $t_{\text{trans}}$ indicate the duration of the superwind phase and the duration of the transition from the AGB wind to superwind phase. The longer $t_{trans}$ the smoother the transition. The tangential variation is determined by the quantities $\alpha$ and $\beta$. $\alpha$ determines the ratio between the density at the equator and the pole:

$$q \equiv 1/(1 - A) \qquad (4)$$

$\beta$ determines the way the density varies from equator to pole. For low $\beta$ values ($< 2$) the density is only high near the equator (flat disk), for high $\beta$ ($> 5$) the density is high everywhere except near the pole (thick disk). See also Mellema et al. (1991), Sect. 3.2.

### 3.3 The fast wind

I call the mass loss phase after the AGB the 'fast wind' even though it is initially only marginally faster than the slow wind. This fast wind supplies the radial inner boundary condition. The first row of cells on the grid is filled with the fast wind and this drives the evolution of the flow. The fast wind is assumed to start accelerating directly after the AGB. Observationally and theoretically the situation is very unclear. Stars on the AGB are thought to lose mass through the process of radiation pressure on dust (see e.g. Bowen & Willson 1991). The hot central stars of PNe are supposed to lose mass through radiation pressure on lines (Pauldrach et al. 1988). It is not very well known under what conditions the first process stops and the second one starts. Consequently the character of the mass loss of stars with an effective temperature between 5000 and 25 000 K is not known. Work of Abbott (1982) indicates that radiation pressure on lines may sustain mass loss in stars with an effective temperature higher than 10 000 K.

Paper III has shown that when the mass loss rate is too low at the time of ionization of the surrounding material, this material collapses back to the centre. Since even young PNe do not typically show a collapsed nebula, a dynamically important wind should be present at the time ionization starts ($T_{eff} \sim 10^4$ K). Also, fast outflows are observed in some transition objects (CRL 618, CRL 2688, OH 231.8+4.2). So, there is some evidence for mass loss between the AGB and PN phases.

As in Paper III I use Kudritzki's algorithm to determine the fast wind properties as a function of the stellar parameters (Kudritzki et al. 1989). This algorithm uses the theory of radiation pressure on lines (CAK theory) and was developed for stars with an effective temperature above 30 000 K. Therefore one could question its application to stars of lower temperature. The velocities derived from the algorithm are of the order of the escape velocity at the surface of the star, and should be roughly correct. It is mainly the mass loss rate that is uncertain. Typical mass loss rates produced by the algorithm for standard parameters are $10^{-7} - 10^{-8}$ $M_\odot$ yr$^{-1}$, much less than on the AGB and therefore not in conflict with observations.

The main effect of using the Kudritzki values for the fast wind for $T_{eff} < 25\,000$ K is that the slow wind is stopped from diffusing inward before ionization and rapidly backfilling after ionization. See also Paper III, Sect. 2.2 for more discussion on the time-dependence of the fast wind.

### 3.4 The central star

For the central star I use the stellar evolution calculations from Schönberner (1983) and Blöcker & Schönberner (1991).

The stellar spectrum is approximated by a blackbody spectrum because model spectra are not available for the entire evolutionary range.

As was discussed in Paper III, Sect. 2.3, these evolutionary tracks are known to have problems. In fact the character of the newer Blöcker & Schönberner tracks is different from the older Schönberner tracks. The older tracks start with a 1000 year period of little or no evolution before the effective temperature starts increasing, the new tracks do not. To correct for this Marten & Schönberner (1991) left out the first 1000 years of evolution when using the old tracks in their (one-dimensional) gasdynamic calculations for PN formation. I do the same here.

Since these calculations are the only ones readily available, I will use them. They reproduce the main characteristics of post-AGB evolution, only the derived time scales are uncertain.

## 4 GASDYNAMIC EVOLUTION

Using the initial conditions described above, several simulations were run. Because of the small grid the simulations only show the early evolution of aspherical PNe. Future work will address the later stages of nebular evolution.

This section is dedicated to the discussion of the two new gasdynamic aspects introduced by using a evolving instead of a constant environment. The first of these is the modification of the slow wind density distribution by the ionization and the accompanying shock fronts. Some of this was already seen in Paper III, where the ionization front was shown to produce surrounding envelopes.

The modification of the slow wind density distribution by the ionization is important for the aspherical interacting winds problem because the shape of the swept-up shell is mainly determined by the ratio of inner pressure over outer density (see e.g. Icke 1988).

The second new aspect is the acceleration of the fast wind. This is expected to introduce a transition from momentum- to energy-driven flow and Rayleigh-Taylor instabilities.

To illustrate these points I use four runs (A to C) whose initial conditions are listed in Table 1. Run D is only used in Sect. 5. I list its parameters here for completeness.

### 4.1 Ionization of the slow wind

For illustrating the effects of the gradual ionization of the slow wind, I make use of the results from runs A and B. These are discussed in the following two subsections.

#### 4.1.1 Run A

Figures 1 and 2 illustrate the evolution of the nebula for run A. This run uses the 0.598 $M_\odot$ Schönberner track and consequently the evolution is relatively slow. The initial density ratio $q = 5.0$ and the shape parameter $\beta = 1.0$, corresponding to a flat initial slow wind density distribution. The superwind is supposed to have lasted 1000 years, so the superwind region extends out to $r = 6.6 \; 10^{14}$ m.

The grey scales in Figs. 1 and 2 show the mass density and the logarithm of the temperature, respectively. The black line in both figures is the contour of 90% H ionization.

**Table 1.** Parameters for the runs presented in this paper.

| Run | A | B | C | D |
|---|---|---|---|---|
| $\dot{M}_{super}$ ( $M_\odot$ yr$^{-1}$) | $2\,10^{-5}$ | $1\,10^{-5}$ | $1\,10^{-5}$ | $1\,10^{-5}$ |
| $\dot{M}_{AGB}$ ( $M_\odot$ yr$^{-1}$) | $6\,10^{-7}$ | $6\,10^{-7}$ | $6\,10^{-7}$ | $6\,10^{-6}$ |
| $t_{super}$ (years) | 1000 | 10000 | 10000 | 10000 |
| $t_{trans}$ (years) | 1000 | 650 | 650 | 650 |
| $\alpha$ | 0.8 | 0.7 | 0.7 | 0.7 |
| $\beta$ | 1.0 | 6.0 | 0.5 | 1.0 |
| $v_0$ (m s$^{-1}$) | $1.2\,10^4$ | $1.5\,10^4$ | $1.5\,10^4$ | $1.0\,10^4$ |
| $T_0$ (K) | $2.0\,10^2$ | $2.0\,10^2$ | $2.0\,10^2$ | $2.0\,10^2$ |
| stellar track ( $M_\odot$) | 0.598 | 0.605 | 0.605 | 0.644 |
| $\dot{M}_{fast}$ ( $M_\odot$ yr$^{-1}$) | $5.0\,10^{-8}$ | $1.0\,10^{-7}$ | $3.6\,10^{-7}$ | $1.1\,10^{-7}$ |
| $k_{fast}$ | 0.055 | 0.053 | 0.053 | 0.070 |
| $r_0$ (m) | $3.0\,10^{14}$ | $1.0\,10^{14}$ | $1.0\,10^{14}$ | $3.0\,10^{14}$ |
| $\Delta r$ (m) | $3.0\,10^{13}$ | $3.0\,10^{13}$ | $3.0\,10^{13}$ | $3.0\,10^{13}$ |
| grid dimension | $70 \times 70$ | $80 \times 80$ | $80 \times 80$ | $70 \times 70$ |

Note: mass loss rates are for the polar direction

This is effectively the ionization front. The six frames show the situation at six consecutive times. At $t = 475$ years the central star has an effective temperature of $10\,10^3$ K and the fast wind has a mass loss rate $\dot{M}_{fw} = 2.9\,10^{-8}$ $M_\odot$ yr$^{-1}$ and velocity $v_{fw} = 210$ km s$^{-1}$. An ionization front has moved into the slow wind and is now changing from an R-front into a D-front. The mild fast wind has kept the central region evacuated. The edge of the density distribution (at $r = 0.035$ pc) corresponds to transition from the superwind to the AGB wind. In the temperature plot the emergence of the hot bubble can be seen ($T \sim 6\,10^5$ K), as well as the temperature increase caused by the ionization front ($T \sim 5\,10^3$ K). As was found in the one-dimensional case (Paper III) the transition from momentum- to energy-driven flow occurs at a fast wind velocity of about 150 km s$^{-1}$. This is in perfect agreement with the analytical results of Kahn & Breitschwerdt (1990).

At $t = 951$ years the ionization front has turned into a full D-front with the corresponding shock wave preceding it. I will call this shock the I-shock (I for ionization) to distinguish it from the shock driven by the bubble of hot shocked fast wind material, the W-shock (W for wind). Likewise the shells swept-up by these shocks are called I- and W-shells.

The I-shock is slightly elliptical, following the elliptical ionization front caused by the aspherical density distribution in slow wind. At this stage the star has $T_{eff} = 15\,10^3$ K and the fast wind $\dot{M}_{fw} = 2.3\,10^{-8}$ $M_\odot$ yr$^{-1}$, $v_{fw} = 326$ km s$^{-1}$. The density jump across the I-shock is almost a factor 2. The D-front constitutes a rarefaction wave, and therefore the velocities increase when going outward from the W-shell to the I-shell. This is basic D-front behaviour, see for instance Lasker (1966), Marten & Schönberner (1991) and Paper III. The temperature in the hot bubble has increased to $2\,10^6$ K, in the ionized and shocked slow wind the temperature is about $6\,10^3$ K.

The next frame in Figs. 1 and 2 shows the situation for $t = 1425$ years. In the polar direction the ionization front has reached the edge of the superwind area and has broken out into the lower density AGB wind region. The I-shell is expanding quickly into this low density area, thereby losing its shell-like appearance. In the equatorial region the nebula is still ionization bounded. The arc in the ionization contour is the edge of the computational grid. The star and fast wind now have $T_{eff} = 21\,10^3$ K, $\dot{M}_{fw} = 2.1\,10^{-8}$ $M_\odot$ yr$^{-1}$, and

$v_{fw} = 476$ km s$^{-1}$. The density and temperature plots show that the hot bubble has acquired an aspherical shape. The W-shell is still of relatively low density ($n \sim 5\,10^9$ m$^{-3}$) compared to the I-shell ($n \sim 2\,10^{10}$ m$^{-3}$). Consequently the W-shell does not feature prominently in the grey scales. The hot bubble has reached a temperature of $8\,10^6$ K, the ionized material is at $7\,10^3$ K.

This picture does not change much for the next frame ($t = 1901$ years). The ionization shock front has almost reached the edge of the superwind region and will break out shortly afterward. The I-shell near the pole has entirely diffused away into the low density AGB wind region.

The fifth frame (at $t = 2376$ years) shows that the ionization front has also reached the edge of the superwind region near the equator and has broken out. The high density I-shell is now emptying itself into the low density AGB wind region, just as it did earlier on near the pole. The star is at $33\,10^4$ K and $\dot{M}_{fw} = 1.7\,10^{-8}$ $M_\odot$ yr$^{-1}$, $v_{fw} = 818$ km s$^{-1}$. The hot bubble temperature is $1.2\,10^7$ K, the ionized slow wind material temperature is $7\,10^3$ K. The asphericity of the hot bubble and the W-shell have increased.

The last frame ($t = 2851$ years) clearly shows the W-shell. Number densities in it range from $3.5\,10^9$ to $5.5\,10^9$ m$^{-3}$. The W-shock is now fully isothermal along the whole shell ($T = 7\,10^3$ K). It is not very strong, compression ratios across it range from 2 to 3. The reason for this is that the pressure of the hot bubble is still relatively low, since the fast wind has not yet reached full speed ($\dot{M}_{fw} = 1.5\,10^{-8}$ $M_\odot$ yr$^{-1}$, $v_{fw} = 1000$ km s$^{-1}$), see also Paper V. Near the symmetry axis the shock may not be fully resolved, because there the shell extends only over two grid cells. This may mean that the actual shell density at that position should be higher. Near the equator the shell is resolved. Notice how the former I-shell has almost completely expanded into the AGB region and how its leading edge has travelled off the computational grid.

This run shows that, as in the one-dimensional case, the early evolution of PN formation is dominated by the ionization front. The photons exert their influence before the fast wind can. In the one-dimensional case the action of the ionization front was shown to result in surrounding envelopes, as will be shown in Sect. 5 this is also true for the two-dimensional case.

Notice the small deformations of the contact discontinuity at times 2376 and 2852 years. These indicate a Rayleigh-Taylor instability, to be discussed in Sect. 4.2. Here I want to concentrate on the influence of the ionization process on the interacting winds mechanism.

The ionization front moves through the aspherical slow wind and we have seen that the radial density distribution is changed because of this. The tangential distribution is important for the formation of the aspherical nebula, and hence it is interesting to see how this is modified by the ionization front. In run A the shape of the bubble is what one would expect from the initial condition ($\beta = 1.0$). Looking in more detail at the density of the ionized slow wind material being swept up shows that the tangential variation indeed stays close to the one in the initial condition. In other words, the 'effective $\beta$' after ionization is close to the original value.

The rate at which the nebula becomes aspherical is determined by the 'effective $\alpha$'. This number is less easily determined. In the early phases the W-shock position

fig/runAd5.ps

fig/runAd10.ps

fig/runAd15.ps

fig/runAd20.ps

fig/runAd25.ps

fig/runAd30.ps

**Figure 1.** Mass density plots for run A. The black line indicates the 90% H$^+$ contour. The extremes are: $8.7\,10^{-21} < \rho(t = 475 \text{ yrs}) < 1.5\,10^{-16}$ kg m$^{-3}$, $4.4\,10^{-21} < \rho(t = 951 \text{ yrs}) < 1.0\,10^{-16}$ kg m$^{-3}$, $2.7\,10^{-21} < \rho(t = 1426 \text{ yrs}) < 7.0\,10^{-17}$ kg m$^{-3}$, $1.8\,10^{-21} < \rho(t = 1901 \text{ yrs}) < 3.7\,10^{-17}$ kg m$^{-3}$, $1.3\,10^{-21} < \rho(t = 2376 \text{ yrs}) < 8.5\,10^{-18}$ kg m$^{-3}$, $9.3\,10^{-22} < \rho(t = 2852 \text{ yrs}) < 9.1\,10^{-18}$ kg m$^{-3}$.

near the equator almost does not change. Notice that between $t = 951$ years and $t = 1901$ years the W-shock at the equator is almost stationary, whereas at the pole the W-shock position increases from 0.017 to 0.025 pc. This is caused by the following effect. The hot bubble is isobaric at a pressure $p_{hb}$ (proportional to $\rho_{fw}v_{fw}^2$ of the fast wind). The ionization of the slow wind leads to a temperature and hence pressure increase of the slow wind ($p_{sw}$), and because

fig/runAt5.ps

fig/runAt10.ps

fig/runAt15.ps

fig/runAt20.ps

fig/runAt25.ps

fig/runAt30.ps

**Figure 2.** Plots of the logarithm of the temperature for run A. The black line indicates the 90% $H^+$ contour. The extremes are: $22 < T(t = 475 \text{ yrs}) < 5.9\,10^5$ K, $14 < T(t = 951 \text{ yrs}) < 2.0\,10^6$ K, $10 < T(t = 1426 \text{ yrs}) < 4.6\,10^6$ K, $10 < T(t = 1901 \text{ yrs}) < 8.3\,10^6$ K, $10 < T(t = 2376 \text{ yrs}) < 1.2\,10^7$ K, $10 < T(t = 2852 \text{ yrs}) < 1.7\,10^7$ K. The very low temperatures occur in pressure undershoots at the inner shock.

$p \propto \rho T$ this pressure increase is higher near the equator. In these early phases $p_{sw}(\theta = \pi/2) \approx p_{hb}$ and therefore the equatorial expansion of the W-shock stalls. At the pole $p_{sw}(\theta = 0) \approx 0.2 p_{sw}(\theta = \pi/2) < p_{hb}$ and the W-shock continues to expand. The same stalling was found in the one-dimensional simulations in Paper III and in the analytical models of Breitschwerdt & Kahn (1990). In this two-dimensional simulation the stalling of equatorial shock helps

fig/sw.modif.ps

**Figure 3.** Density profiles for run A at $t = 2852$ years. The sold line shows the density along the polar axis ($\theta = 0$), the dashed line at the equator ($\theta = \pi/2$). Note the swept-up shell and the difference in radial fall-off in both cases.

creating an aspherical bubble in the early phases.

At the later stages when $\rho_{fw} v_{fw}^2$ and hence $p_{hb}$ have increased, $p_{hb} > p_{sw}$ at all positions. Now it is mainly the density distribution in the slow wind that determines the expansion of the W-shock, since for $p_{hb} \gg p_{sw}$, $v_{shock} \propto \sqrt{p_{hb}/\rho_{sw}}$. Because of the differential blow-out of the I-shock into the AGB-wind region, $\rho_{sw}$ is a rather complex function of position. Consider the situation at $t = 2852$ years (Fig. 3). Just in front of the shock the density contrast in the slow wind is only 2.5 (compared to 5.0 in the initial conditions). But the density along the polar axis falls off as a steep function of $r$ (approximately $r^{-3.5}$), whereas the density along the equator is almost constant. So, at larger $r$, the density contrast is larger (around 4.0). This situation has arisen because in the polar direction the ionization shock front and its accompanying I-shock have broken out at an earlier stage (around $t = 1200$ years). The matter swept up by the I-shock has expanded into the low density AGB wind, giving rise to the steep radial density profile. In the equatorial region the front has only recently broken out and the radial density profile is still rather shallow. Note that at $t = 2376$ years the density along the equator actually goes up with radius (see Fig. 1).

So, the density contrast just in front of the swept-up shell is low, suggesting a slow increase in asphericity, but the radial density profile along the pole is much steeper than along the equator, suggesting a quick increase in asphericity. Looking at the actual shock velocities shows that at the end of the simulation ($t = 3\,802$ years) the W-shock is still expanding five time faster at the pole than at the equator. For these initial conditions ($\alpha = 0.8$) this ratio in the non-radiative case would have been

$$\left(\frac{v_{shock}(\theta=0)}{v_{shock}(\theta=\pi/2)}\right)_{non-rad} = \frac{\sqrt{p_{hb}/\rho(\theta=0)}}{\sqrt{p_{hb}/\rho(\theta=\pi/2)}} = \sqrt{\frac{1}{1-A}} = \sqrt{5} \approx 2.2 \ . \tag{5}$$

So the modification of the slow wind by the ionization front can lead to a considerable increase in the rate at which the nebula becomes aspherical.

As was pointed out in Paper V, smoothing by pressure waves will diminish the slow wind asphericity within a few sound crossing times (3000 years for a distance of $10^{15}$ m and a temperature of $10^4$ K). In the long run the asphericity of the nebula will therefore stop increasing. This stage is not reached in any of the simulations presented here.

### 4.1.2 Run B

In run B the initial slow wind distribution is a thickened disk, i.e. $\beta$ has a high value, see Table 1. The density contrast is fairly mild, $q = 3.3$. Figure 4 shows the evolution in mass density plots with H ionization overlay, Fig. 5 the same for the logarithm of the temperature. Numerical problems (the time step became too small due to temperature overshoots) stopped the simulation at $t = 1635$ years, before a substantial W-shell formed. I nevertheless present the result here because of the modification of slow wind density distribution it shows.

This simulation shows the same basic behaviour as run A: the H ionization front becomes a D-front with an accompanying shock and this front breaks out quickly in a cone in the polar direction at $t = 1141$ years ($T_{eff} = 18\,000$ K). The rest of the grid follows at $t = 1426$ years ($T_{eff} = 23\,000$ K). This is at a much lower temperature than in run A because the densities in the polar and equatorial direction are lower by factors 2 and 3.3 respectively (see Table 1).

At the end of the simulation the star has a temperature of 26 000 K and the fast wind has a modest $v_{fw} = 600$ km s$^{-1}$ and $\dot{M}_{fw} = 2.0\,10^{-8}$ M$_\odot$ yr$^{-1}$. The W-shell is just starting to reach substantial densities. The shape of the W-shell deviates strongly from what is expected on the basis of the initial conditions and constant environment results from Papers IV and V. The ionization front has modified the density in the slow wind in such a fashion that the shock travels much faster in a very narrow region along the symmetry axis, much narrower than is expected on the basis of the initial value of $\beta$. In other words, the effective $\beta$ has increased because of the modification of the slow wind by the ionization front. For run A no such effect was found. This is not as surprising as it seems, because for a flat distribution as in run A ($\beta = 1.0$) a little increase in the height of the disk does not make much of a difference, but when the original funnel is already quite narrow such as in run B ($\beta = 6.0$), a little increase can have dramatic consequences.

This effect is potentially important for the formation of FLIERS (Fast Low Ionization Regions) often seen along the major axis of PNe. A more thorough investigation of this effect should be part of future work. Note that the reports of anomalous nitrogen abundances in FLIERS seem to imply that some other process beside wind-wind interaction is needed to explain their full structure (Balick et al. 1993).

### 4.2 Instabilities

Because the fast wind velocity increases in time in the evolving environment case, it is to be expected that instabilities of the Rayleigh-Taylor type will form. These instabilities are due to the density inversion in an effective gravity (the outward acceleration of the shell). The analytical work of Kahn & Breitschwerdt (1990) and Breitschwerdt & Kahn (1990) indicated that the growth times are indeed small enough (300 to 600 years) for Rayleigh-Taylor instabilities to form.

fig/runBd5.ps

fig/runBd10.ps

fig/runBd15.ps

fig/runBd20.ps

**Figure 4.** Mass density plots for run B with the black line showing the 90% H$^+$ contour. The extrema are: $1.7\,10^{-20} < \rho(t = 475 \text{ yrs})$ $< 8.4\,10^{-17}$ kg m$^{-3}$, $6.0\,10^{-21} < \rho(t = 951 \text{ yrs}) < 3.4\,10^{-17}$ kg m$^{-3}$, $2.4\,10^{-21} < \rho(t = 1426 \text{ yrs}) < 1.4\,10^{-17}$ kg m$^{-3}$, $2.9\,10^{-30}$ $< \rho(t = 1635 \text{ yrs}) < 1.1\,10^{-17}$ kg m$^{-3}$. The very low density in the last frame is an example of a hot spot that effectively stops the integration.

Instabilities are numerically hard to follow since they physically develop on small length scales. The smallest length scale possible in a simulation is determined by the grid cell size and the development of the instabilities will thus depend on the resolution of the grid (see Fryxell, Müller & Arnett 1991 for an illustration of this effect). The simulations presented here were done on a coarse grid and are therefore not particularly well suited to studying instabilities. This can be seen in the runs presented above. These only show marginal signs of instabilities at the contact discontinuity (note the mild corrugation on the inside of the W-shell in Fig. 1 at times 2376 and 2852 years and in Fig. 4 at 1426 and 1635 years).

A run which does show the Rayleigh-Taylor instability is presented in Fig. 6. This run has a rather mild density contrast between pole and equator ($q = 3.33$) and the slow wind has a flat distribution ($\beta = 0.5$). We see the same type of evolution as in runs B and C. The ionization front sweeps up an I-shell and later the W-shell forms, but here the W-shell is disrupted by a Rayleigh-Taylor instability. Tongues of dense swept-up slow wind material penetrate the hot bubble. Because of the low resolution this result should only be interpreted as a sign that the instability occurs. It should not be used to derive characteristics of the instability.

Cooling shells driven by hot bubbles are subject to other instabilities as well (Vishniac 1983; Bertschinger 1986), and it cannot be ruled out that the instabilities seen here are actually of this type. Their scale length would favour this explanation, but since they are not seen in all simulations, it remains hard to state something definite about their nature.

fig/runBt5.ps

fig/runBt10.ps

fig/runBt15.ps

fig/runBt20.ps

**Figure 5.** Plots of the logarithm of the temperature for run B with the black line showing the 90% H$^+$ contour. The extrema are: $24 < T(t = 475 \text{ yrs}) < 6.8\,10^5$ K, $24 < T(t = 951 \text{ yrs}) < 2.9\,10^6$ K, $22 < T(t = 1426 \text{ yrs}) < 5.1\,10^6$ K, $10 < T(t = 1635 \text{ yrs}) < 1.1\,10^{16}$ K. The very high temperature in the last frame is an example of a hot spot that effectively stops the integration.

Breitschwerdt & Kahn (1990) also predicted Rayleigh-Taylor instabilities in the I-shell. I find marginal indications for this in one simulation, but as was said above, these simulations are not really suited for a proper stability analysis.

### 4.3   Time scales

Finally I want to point to an important result from these simulations, namely that the aspherical interacting winds scheme can indeed shape aspherical nebulae on the time scales needed. Previous models (Papers IV and V) used a constant fast wind and the typical ages of the nebulae obtained were around 1000 years. Here the formation of a PN is followed using published stellar evolution tracks and mass loss prescriptions. After 3000 years of post-AGB evolution the nebula has reached a major axis diameter of 0.1 pc and a minor axis diameter of 0.05 pc; hence a fairly aspherical nebula with a reasonable size develops in an early phase.

In fact, even a somewhat slower evolution would be reasonable. A slower evolution is expected when the fast wind starts later or the momentum flux in the fast wind is lower than assumed here. The aspherical interacting winds model can thus easily produce aspherical nebulae on the right time scale. Note also that the models seem to indicate that initial density contrasts of 3 to 5 suffice to create aspherical nebulae in the time available.

### 5   OBSERVABLES

In this section I describe the observables derived from the simulations. The observables are emission line images and kinematic data (long slit spectra). After discussing them I will compare them to observations. Since we are dealing with young objects, the observational material is a poor constraint on the models. Most young nebulae are small (compared to the resolution of images made of them) and appear structureless. In addition there seems to be a selection effect against observing very young PNe: only PNe with central stars of 25 000 K and higher are known. Presumably very young objects are obscured by dust (see Pottasch 1993).

fig/runCd5.ps

fig/runCd10.ps

fig/runCd15.ps

fig/runCd20.ps

fig/runCd25.ps

**Figure 6.** Mass density plots for run B. The black line shows the 90% $H^+$ contour. The extrema are: $1.9 10^{-20} < \rho(t = 475 \text{ yrs}) < 8.4 10^{-17}$ kg m$^{-3}$, $3.7 10^{-21} < \rho(t = 951 \text{ yrs}) < 3.4 10^{-17}$ kg m$^{-3}$, $1.3 10^{-21} < \rho(t = 1426 \text{ yrs}) < 1.4 10^{-17}$ kg m$^{-3}$, $1.4 10^{-22} < \rho(t = 1901 \text{ yrs}) < 7.5 10^{-18}$ kg m$^{-3}$, $1.8 10^{-22} < \rho(t = 2376 \text{ yrs}) < 4.5 10^{-18}$ kg m$^{-3}$.

### 5.1    Images from the simulations

Figure 7 shows $H\alpha$ images seen at $0°$ inclination in the sky for run A at the six times shown in Figs. 1 and 2. These images are constructed by taking the two-dimensional emissivity data from the simulation, rotating them around the symmetry axis to obtain the three-dimensional picture and projecting this on the sky at an inclination $i$. Because of the rotation any bright spots in the two-dimensional emissivity data will appear as stripes (for $i = 0°$) or ellipses ($i \neq 0°$) in the projected images. This should be kept in mind when interpreting these images.

The figure shows that the images of the early stages mainly show the ionization front ($t = 475, 951, 1426, 1901$ years), I will call this the 'ionization phase'. Because of the asphericity of the slow wind, these images show an elliptical or barrel shape. At $t = 2376$ the inner elliptical rim (the W-shell) appears. It is surrounded by a bright envelope, which by $t = 2852$ years has become much fainter. However, one should be cautious in interpreting the last image since at that stage some of the higher density ionized slow wind material has flowed off the grid and is no longer seen in the synthesized image. I will call this second phase the 'rim-envelope phase'.

Figure 8 shows [NII]-6384Å images and Fig. 9 [OIII]-5007Å images for run A (henceforth referred to as [NII] and [OIII]). The [NII] images follow the Hα images so at $t = 2376$ years the W-shell is not seen in [NII]. This is because of the low N$^+$ density in the shell: most of the N there is N$^{2+}$. At $t = 2852$ years the shell shows up, albeit fainter than the envelope. The [OIII] images for $t = 475$ and 951 years are low intensity noise, only from $t = 1426$ years onward a real image is found. Because O$^{2+}$ is abundant in the inner regions, the swept-up shell is seen in all [OIII] images. Note that times $t = 1426$ and 1901 years we see an entirely different nebula in [OIII] than in [NII] and Hα.

This type of evolution is seen in all simulations. Initially there is an ionization phase in which an aspherical ionization bounded nebula is seen. After the nebula has become density bounded, the W-shell starts showing up as a bright rim, surrounded by an envelope of ionized slow wind gas, the rim-envelope phase.

Another example is shown in Fig. 10. This figure contains Hα images from run D at $t = 1426$ years, seen at four different inclinations in the sky. In run D the star has evolved quicker (it follows the 0.644 M$_\odot$ Schönberner track, see Schönberner 1983) and the superwind region extends beyond the edge of the grid; therefore no blow-out of the I-shell into the low density AGB wind region has occurred and the I-shell has not yet flowed off the grid. It shows up as the sharp edge of the envelope at $r = 0.04$ pc. On the inside the W-shell shows up as a bright rim. This nebula is therefore in the rim-envelope phase.

## 5.2 Kinematic data from the simulations

Figures 11 and 12 show the [NII] and [OIII] long slit spectra (position-velocity diagrams) taken along the major axis of the model nebulae placed at an inclination of $0°$ in the sky. Because there is virtually no [OIII] emission in the early stages (see Fig. 9), only the diagrams for $t = 1426$ years and later are shown. The [NII] spectra show how the ionized material slowly accelerates from an expansion velocity of slightly less than 20 km s$^{-1}$ at $t = 475$ years to 25 km s$^{-1}$ at $t = 1901$ years. This acceleration is caused by the gradual ionization of the I-shell, which (because it forms a rar-efaction wave) has higher velocities further out. When the ionization front reaches the edge of the superwind, it breaks out and the ionized gas accelerates up to velocities of almost 50 km s$^{-1}$. This is seen at $t = 2376$ years. At $t = 2852$ years some of this high velocity gas has flowed off the grid and therefore no longer shows up in the synthesized spectrum. At that time the W-shell starts showing up in the [NII] spectrum. The bow-tie shape of the shell in the diagram is caused by the slow expansion at the equator and a larger (sideways) expansion near the poles.

The [OIII] position-velocity diagrams display the evolution of W-shell swept up by the hot bubble. Its expansion velocity in these early stages is very low (only a few km s$^{-1}$) and therefore the splitting of the line is not resolved. These low expansion velocities are related to the stalling of the W-shell expansion near the equator described in Sect. 4.1.1. The pressure increase due to the ionization of the slow wind keeps the W-shock from expanding outward. Only at $t = 2852$ years the fast wind velocity and hence the hot bubble pressure have acquired a high enough value to cause a substantial expansion of the W-shell near the equator. As in the [NII] a bow-tie shape is seen, indicating rapidly expanding polar lobes. Notice that the [NII] velocities are typically higher than the [OIII] velocities. This is caused by the rarefaction wave pattern in the velocity (see Sect. 4.1.1). Behaviour like this is seen in many nebulae and was once thought to indicate that the nebulae are simple shells expanding into near vacuum (see e.g. Osterbrock 1989., p. 189 – 197).

Figure 13 shows the major axis long slit spectra corresponding to the images from run D in Fig. 10. Because of the faster evolution in this case, the bright rim is expanding at about the same velocity as the envelope near the equator and even faster at the poles. The spectra show velocities of about 25 km s$^{-1}$ at $i = 5°$ and 35 km s$^{-1}$ at $i = 80°$. Notice how little the envelope velocity varies with inclination, it expands uniformly in all directions with a velocity of about 25 km s$^{-1}$. Compare these figures with observed slit spectra in Sabbadin, Bianchini & Hamzaoglu (1984), Chu (1989), O'Dell, Weiner & Chu (1990).

Similar behaviour was found in one-dimensional simulations (Paper III), where it was shown that the rarefaction wave set up by the ionization front can produce nebulae observed to be expanding slower than the surrounding envelope.

## 5.3 The ionization phase

The results discussed above show that there are two ways of producing an aspherical nebula. In the early phases of PN formation aspherical morphologies occur because of the aspherical ionization front and at later times because of the wind-wind interaction. The early aspherical nebulae look like some nebulae formed by wind-wind interaction: elliptical and opened-up elliptical (barrel-shaped) nebulae are found. One may wonder whether there is an observational test by which to distinguish the two. This is important since otherwise we have two ways to explain a given nebula.

These ionization shaped nebula are found only to occur in the early phases of PN evolution. An observational test would thus be the occurrence of ionization fronts or a clearly different morphology in [OIII] such as shown by the images

fig/runAimrh5.ps

fig/runAimrh10.ps

fig/runAimrh15.ps

fig/runAimrh20.ps

fig/runAimrh25.ps

fig/runAimrh30.ps

**Figure 7.** H$\alpha$ images of run A seen at $i = 0°$ in the sky.

fig/runAimn1a5.ps

fig/runAimn1a10.ps

fig/runAimn1a15.ps

fig/runAimn1a20.ps

fig/runAimn1a25.ps

fig/runAimn1a30.ps

**Figure 8.** [NII] images from run A seen at $i = 0°$ in the sky.

fig/runAimo2a5.ps                    fig/runAimo2a10.ps

fig/runAimo2a15.ps                   fig/runAimo2a20.ps

fig/runAimo2a25.ps                   fig/runAimo2a30.ps

**Figure 9.** [OIII] images from run A seen at $i = 0°$ in the sky.

fig/runDha5.ps

fig/runDha30.ps

fig/runDha55.ps

fig/runDha80.ps

**Figure 10.** Hα images from run D at $t = 1426$ years shown at four different inclinations. The star (0.644 M$_\odot$) has an effective temperature of 65 000 K.

in Figs. 7 – 9. It may be that some nebulae stay optically thick for a longer time than seen in these simulations; in that case this criterion would not be very good.

The only other clear difference between the two types is the kinematic behaviour. Wind blown bubbles show velocity differences between the different parts of the nebula: the expansion along the major axis is faster than along the equator. Ionization driven bubbles show virtually no velocity variation. The reason for this is that a discontinuity set up by a weak ionization D-front travels approximately with the speed of sound (see e.g. Shu 1992, p. 278 – 282). This means that its velocity is almost independent of the outer density. In a wind-wind interaction high shock speeds and shell velocity variations of the order of the outer density variation can occur (see also Paper IV).

Since only the projected velocities can be measured it may sometimes prove difficult to make the distinction, also because real PNe are likely to be more complex than the models. However, all modelling to date suggest that the ionization phase is relatively short and consequently only very young and small nebulae are likely candidates for this. In the sample of Balick (1987) the only serious candidates are IC 418 and BD 30°+3639. I do not know of any detailed kinematic data to check their evolutionary state.

Another interesting object in which ionization may play a role in the shaping is NGC 40. In Hα and [NII] it shows a barrel-shaped nebula (an opened-up ellipse). In [OIII] it shows a closed elliptical rim, see Fig. 14. This suggests an ionization front for O$^{2+}$. It may also be that this nebula is in the phase in which we see in Hα and [NII] the I-shell and in [OIII] the W-shell (cf. Figs. 7 – 9, $t = 1426$ years). This would explain why the morphology in Hα is very different from that in [OIII]. Hα and [NII] emission can be seen outside of the main nebula, suggesting that the nebula is no longer entirely ionization bounded.

Pleading against the above interpretation is the fact that the [OIII] rim is only marginally separated from the Hα/[NII] rim. It may be that we see the bright rim overtaking the ionization rim. Another disquieting fact is that Bianchi & Grewing (1987) find that the UV spectrum of the central star indicates a temperature of 90 000 K. This is in contrast with the value of 30 000 K derived using nebular methods (see e.g. Kaler & Jacoby 1991). Bianchi & Grewing envisage a 'curtain' of carbon around the star to soften the spectrum. This would explain why the surrounding nebula suggests a much lower stellar temperature. If this is indeed the case, the evolution of the UV flux and hence of the nebular ionization structure may have been very different from the simulations, since the star need not have followed a regular stellar evolution track.

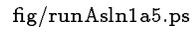

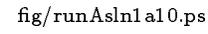

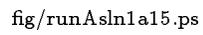

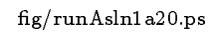

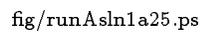

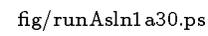

**Figure 11.** Long slit [NII] spectra of run A along the major axis at $0°$ inclination in the sky. The diagrams have been convolved with a Gaussian of 6  $km\,s^{-1}$ FWHM.

fig/runAslo2a15.ps

fig/runAslo2a20.ps

fig/runAslo2a25.ps

fig/runAslo2a30.ps

**Figure 12.** Long slit [OIII] spectra of run A along the major axis at 0° inclination in the sky. The diagrams have been convolved with a Gaussian of 6 km s⁻¹ FWHM

Balick et al. (1987a) suggested that the [OIII] originates from a thin interface between the W-shell and the hot bubble of shocked fast wind material and the [NII] and Hα emission from the main part of the W-shell. The fact that the [OIII] rim lies close to the [NII] rim supports this view. As was shown in Paper IV the interacting winds mechanism can indeed form barrel-shaped nebulae such as NGC 40. However, in this explanation it is not entirely clear why the [OIII] morphology should differ from that of the Hα and [NII]. If the [OIII] emitting gas is located at the interface of the W-shell and the hot bubble, it should show the same shape as the rim, since the W-shell is expected to be quite thin (cf. Mellema et al. (1991) and Paper IV).

An observational test to distinguish the two explanations is hard to find. In both suggestions the [OIII] rim is expected to have a lower density and a higher temperature than the Hα and [NII] rim. Observations by Clegg et al. (1983) show that the temperature derived from [OIII] is indeed some 3 000 K higher than seen in [NII]. In the Balick et al. explanation the [OIII] must have a higher expansion velocity than the Hα and [NII], and this is indeed observed: Acker et al. (1992) list a value of 29 km s⁻¹ for [OIII] and 26 km s⁻¹ for [NII]. In the two-shell explanation the velocity of the inner [OIII] shell can be both higher and lower than that of the Hα and [NII] rim, depending on the characteristics of the fast wind. If the inner rim is indeed overtaking the outer rim, its velocity is expected to be higher, in line with what is observed. The [NII] kinematic data presented in Balick, Preston & Icke (1987b) show a constant expansion velocity of about 25 km s⁻¹. This is what is expected for a I-shell, but because the symmetry axis of NGC 40 lies almost in the plane of the sky ($i \approx 0°$) any polar expansion is hard to observe. A set of long slit spectra of the [OIII] rim at taken at different positions might shed more light on the situation. If this shows more kinematical variation than the [NII], this would support the two-shell picture.

fig/runDhasl5.ps

fig/runDhasl30.ps

fig/runDhasl55.ps

fig/runDhasl80.ps

**Figure 13.** Long slit H$\alpha$ spectra of run D along the major axis at $t = 1426$ years for four different inclinations. The diagrams have been convolved with a Gaussian of 6 km s$^{-1}$ FWHM.

### 5.4   The rim-envelope phase

After the hot bubble has swept up a substantial W-shell, the nebula is observed to consist of a bright inner rim and a surrounding envelope. This phase can be seen in many PNe. Chu, jacoby & Arendt (1987) estimate that at least 50% of all PNe are surrounded by envelopes. The observed envelopes are typically less than three times larger than the inner rim and are diffuse. They often show sharp edges but are seldom edge-brightened. They are commonly identified with the last high mass loss phase on the AGB (the superwind) (see e.g. Chu 1989; Frank, Balick & Riley 1990). However, the results in Paper III as well as the results presented here, show that they acquire their characteristic shape through the ionization front dynamics and they therefore contain only indirect information on the AGB mass loss history. Some authors call these envelopes "attached shells" or "inner haloes". I use the term envelope and reserve 'shell' to indicate wind-swept shells.

Nebulae with clear envelopes in the sample of Balick (1987) are NGC 1535, 2022, 2392, 2610, 3242, 6826, 6894, 7009, 7354, 7662, IC 289, 1454, 3568, and He 1-4. This is about one quarter of the sample. Figure 15 shows some examples. These observed envelopes often show structure. Their sharp edges can be elliptical and the intensity may vary tangentially. In the nebulae mentioned above the major over minor axis ratio varies from 1.0 to 1.3. The same ratio for the inner bright rim is typically 0.1 larger, i.e. the inner rim is more elliptical than the envelope (see also Balick et al. 1992). Almost always the envelope is brighter along the minor axis. At the same distance from the central star the major axis intensity can be 20 to 70% lower than the minor axis intensity. Often the contrast is larger near the edge of the envelope. This can be understood from projection effects. The presence of knots of high intensity along the major axis in some of these objects makes it difficult to obtain these numbers and they should therefore be considered as estimates.

fig/n40imha.ps                                    fig/n40imo2.ps

Figure 14. Images of NGC 40 in Hα and [OIII]. Note the different morphology of the [OIII] image. Images from Balick (1987).

If these envelopes are the undisturbed remnants of a superwind phase, their asphericity calls for a tangential velocity variation in the superwind. This cannot be ruled out, but the simulations in this paper show that if the envelopes are formed by ionization dynamics, the density contrast in the slow wind will naturally produce envelopes with aspherical edges and intensity variations. The ellipticities of the modelled envelopes are not as extreme as the observed ones. The major to minor axis ratios range from 1.0 to 1.1. However, asphericity is there and at later stages the envelopes may grow more aspherical. Other radial density profiles in the slow wind may also result in higher ellipticities. Note that this explanation for the shapes of envelopes naturally results in the inner bright rims being more aspherical than the envelopes.

The models show that at least at some stages and some inclinations the outer edge of the envelope is brighter than the inside (Fig. 10, $i = 80°$; see also Figs. 9 – 10 in Paper III). Such nebulae are indeed observed. Good examples are IC 418, IC 4637, NGC 6804, and (only along the minor axis) NGC 7009. Some others may be in this stage (NGC 2392, NGC 4361, NGC 7139, NGC 7662). See Balick (1987) and Schwarz, Corradi & Melnick (1992) for images of these nebulae. Edge-brightened envelopes (or envelopes with 'crowns' in the Frank et al. 1990 terminology) seem to be rare. This places constraints on the initial density distribution in the slow wind and hence on the mass loss history of the star.

### 5.5 Kinematics of envelopes

Because envelopes are set up by the I-shock, their expansion velocities are expected to be of the order of the velocity of sound. Somewhat higher velocities may occur when the I-shock encounters steeper than $r^{-2}$ density variations in the slow wind (see run A).

Kinematic data by Chu & Jacoby (private communication) for a sample of seven objects (NGC 1535, 2022, 3242, 6826, 7662, IC 3568, 4593) show that envelope expansion velocities range from 40 to 50 km s$^{-1}$. This is a fairly narrow range. Note that these values alone already indicate that this slow wind material has been disturbed, since typical AGB mass loss velocities range from 5 to 20 km s$^{-1}$. Notice also that the observed velocities are higher than those from run D, suggesting a steeper than $r^{-2}$ slow wind density variation.

The internal kinematic structure of the envelopes has not been studied very well. The kinematic data in Balick et al. (1987b) contain some information, but this is not discussed in detail by the authors. It seems that the envelopes contain no variation in their velocity field: they expand with the same velocity in all directions. Note that often the envelopes are low surface brightness and only show up in the Hα kinematic data. Since these suffer from a large amount of thermal broadening, any kinematic patterns may be smoothed out.

The simulations indicate that envelopes are indeed kinematically boring. When one interprets the envelopes as the remnants of the ionization front this is not surprising.

Since these envelopes are formed of slow wind material and show some very typical morphological and kinematic properties, a further study of the development of these envelopes will certainly help in unravelling some aspects of the mass loss history on the AGB.

### 6 CONCLUSIONS

This paper shows the results of some simulations for the formation of aspherical planetary nebulae. The evolutionary pattern found is rather complex since position- and time-dependent ionization and interacting winds dynamics play a role. The stellar and fast wind evolution were taken from published models. For these the ionization processes modify the state the surrounding material before the fast wind has acquired enough momentum to sweep up a substantial shell into it.

fig/n2022.ha-env1.ps

fig/n3242.ha-env1.ps

fig/n6826.ha-env1.ps

fig/n7354.ha-env1.ps

**Figure 15.** Four images from Balick (1987). The grey scales of Hα images have been adapted to show the structure of the envelopes.

These modifications and their effects are:

1. The ionization raises the pressure of the slow wind material. This pressure increase may stall or slow down the expansion of the hot bubble. Because the ionized slow wind is isothermal but not isobaric and the hot bubble is, the bubble expansion is hindered more near the equator. This leads to an effective increase in the pole to equator density ratio. As the hot bubble pressure rises, this effect becomes less important. Through this mechanism the initial increase in asphericity may be fairly large.

2. The ionization front turns into a D-front with a preceding shock (which I call I-shock to discern it from the shock caused by the wind-wind interaction, the W-shock). The D-front flow pattern modifies the density and velocity distribution of the surrounding material:

a. The I-shock sweeps up a dense shell in the surrounding material. This I-shell shapes an early aspherical nebula, which may be unobservable due to obscuration by dust. This early nebula is seen in later stages as an envelope surrounding the shell swept up by the W-shock. Because of the (assumed) density variations in the surrounding material this early nebula and the later envelope assume an elliptical shape and are brighter along the minor axis.

b. The material inside the D-front shell acquires a radial density profile which is initially increasing outward and whose subsequent evolution depends on the initial density distribution in the slow wind. For instance, a blow-out into a lower density (pre-superwind) region can lead to steep radial density fall-offs ($r^{-4}$).

c. For a disk-like slow wind density distribution, the tangential density distribution is not observed to be substantially modified by the D-front. But when the slow wind density is distributed toroidally, there is some indication that this is amplified by the D-front, i.e. the D-front makes the funnel narrower. This may be an important effect for the formation of FLIERS (Fast Low Ionization Regions).

d. The D-front sets up a rarefaction pattern in the radial velocity. This means that the hot bubble sweeps up a shell in a medium that is accelerating outward and has a low velocity near the swept up shell. This can account for observations in which the expansion velocities in PNe are seen to increase outward.

As was said in the introduction, the computational effort presented in this paper may be somewhat premature since the initial and boundary conditions are so poorly known. Planetary nebulae come in many different shapes and with varying kinematic properties and so do the simulations. In analyzing simulations of this kind one should be careful to concentrate on the general features and their physical background and not lose oneself in the particulars of one specific simulation. When so much is unknown, this becomes increasingly hard. On the basis of this the main conclusion from the work presented in this paper should be that the aspherical interacting winds mechanism which until now had only been explored under generalised conditions, now has been shown to be able to produce aspherical nebulae under realistic conditions. Ionization fronts modify the properties of the surrounding slow wind somewhat, but not enough to change the basic aspherical interacting winds picture. This modification does offer an explanation for the origin of envelopes often seen around PNe.

## ACKNOWLEDGEMENT

I would like to thank Adam Frank, Vincent Icke and Bruce Balick for all the discussions and support during the time we worked together. Thanks to the people of room 624 at Leiden Observatory for putting up with the appalling colours and the odd chair of the 'does'. Rolf Kudritzki was kind enough to give me his program to calculate the fast wind properties from the stellar parameters. Some of the calculations presented in this paper were done on the Amsterdam Cray YMP, operated by the Netherlands Council for Supercomputer Facilities (NCF). My research assistantship at UMIST is PPARC funded.